\documentclass[preprint,prX,nofootinbib]{elsarticle}
\usepackage{amssymb}
\usepackage{multirow}
\usepackage{array}
\usepackage{enumerate}
\usepackage{footnote}
\usepackage{graphicx}
\usepackage{multicol}

\begin{document}

\begin{frontmatter}

\title{Characterization of the Hamamatsu R11410--10 3-Inch Photomultiplier Tube for Liquid Xenon Dark Matter Direct Detection Experiments}
\author[UCLA]{K.~Lung\corref{cor1}}
\ead{kvlung@physics.ucla.edu}

\author[UCLA]{K.~Arisaka}
\author[UCLA]{A.~Bargetzi\fnref{fn1}}

\author[UCLA]{P.~Beltrame}
\author[UCLA]{A.~Cahill}
\author[Hamamatsu]{T.~Genma}
\author[UCLA]{C.~Ghag}
\author[UCLA]{D.~Gordon}
\author[UCLA]{J.~Sainz}
\author[UCLA]{A.~Teymourian}
\author[Hamamatsu]{Y.~Yoshizawa}

\cortext[cor1]{Corresponding author. Tel: +1 (310) 825-1902}
\fntext[fn1]{Current Address: Institute for Particle Physics, ETH Z\"{u}rich, \\CH-8093 Z\"{u}rich, Switzerland}

\address[UCLA]{Department of Physics and Astronomy, University of California, Los Angeles,\\475 Portola Plaza, Los Angles, CA 90095, USA}
\address[Hamamatsu]{Electron Tube Division, Hamamatsu Photonics K.K., 314-5 Shimokanzo, Iwata City 438-0193, Shizuoka, Japan}

\begin{abstract}
To satisfy the requirements of the next generation of dark matter detectors based on the dual phase TPC, Hamamatsu, in close collaboration with UCLA, has developed the R11410-10 photomultipler tube. In this work, we present the detailed tests performed on this device. High QE ($\sim$30\%) accompanied by a low dark count rate (50 Hz at 0.3 PE) and high gain ($1\times10^{7}$) with good single PE resolution have been observed. A comprehensive screening measurement campaign is ongoing while the manufacturer quotes a radioactivity of 20 mBq/PMT. These characteristics show the R11410-10 to be particularly suitable for the forthcoming zero background liquid xenon detectors. 

\end{abstract}


\end{frontmatter}

\section{Introduction}
\label{sec:introduction}
Dual phase xenon time projection chambers are one of the prominent technologies for direct dark matter detection~\cite{Wang,Scint,Xe100Ins,ZEP,LUX,G2G3G4}. Interactions occurring in the active volume produce primary (S1) and proportional scintillation (S2) signals observable by arrays of photomultiplier tubes. The two signals provide energy information and signal/background discrimination. To achieve the best sensitivity, next generation xenon detectors on the ton scale require low energy thresholds (below 10 $keV_{nr}$, nuclear recoil equivalent energy) and low background due to both radioactivity and noise. The Hamamatsu R11410-10 photomultiplier tube has been designed and developed for this purpose. 

This work reports the results of detailed measurements performed at UCLA and Hamamatsu and is organized as follows: Section~\ref{sec:R11410} focuses on the physical characteristics and the material composition of the PMT; Section~\ref{sec:setup} introduces the experimental setup employed at UCLA; Sections~\ref{sec:signal} and~\ref{sec:noise} describe the results of the measurements of signal response and noise properties.

\begin{figure}[htb]
\begin{center}
\includegraphics[height = 55mm]{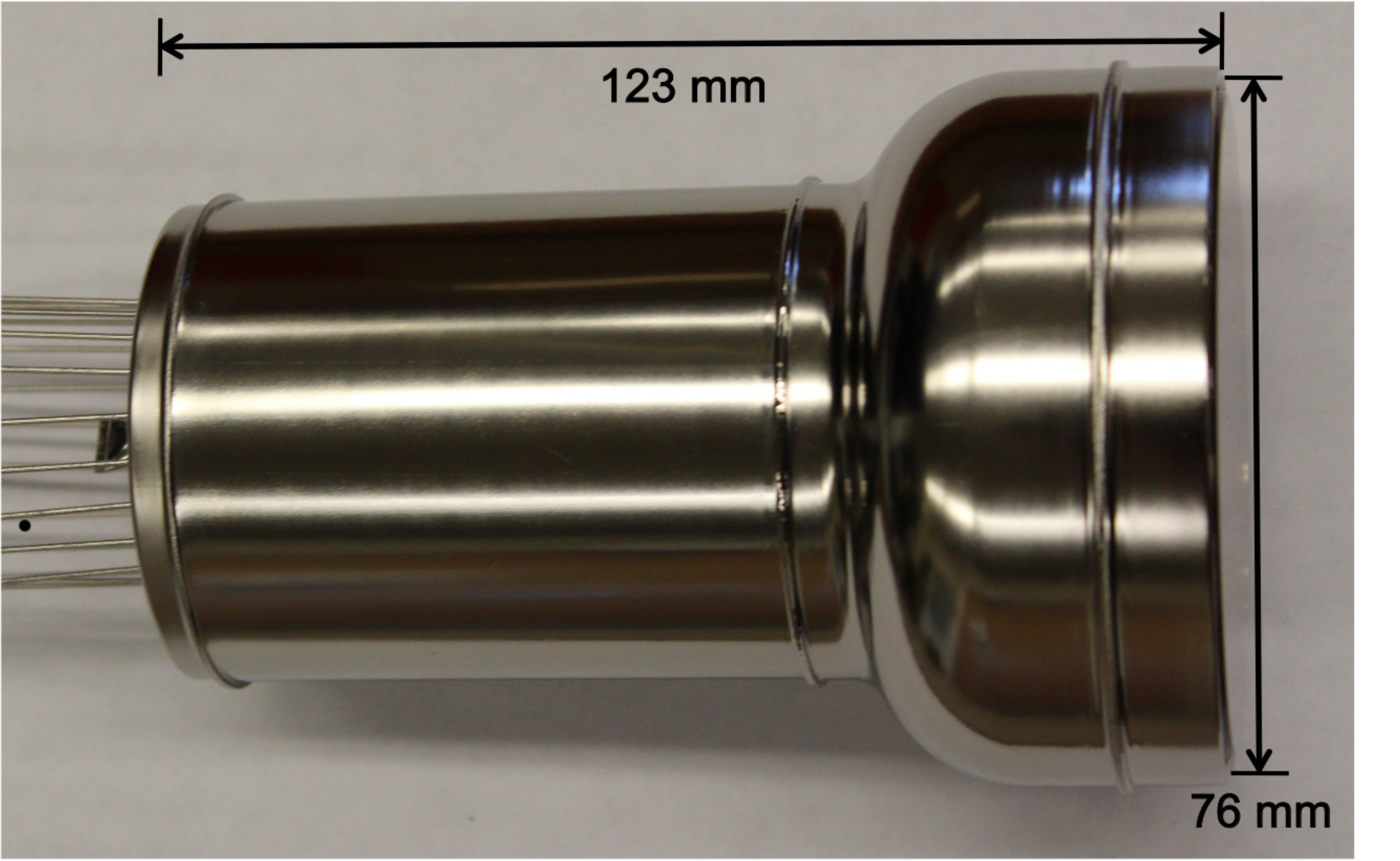}
\end{center}
\caption{
 \label{fig:r11410_picture}
 External structure of the R11410-10 PMT. The physical dimensions are labeled in millimeters.
}
\end{figure}

\section{Hamamatsu R11410--10 3-inch Photomultiplier Tube}
\label{sec:R11410}
The R11410--10 photomultiplier tube is a vacuum device with a transparent synthetic silica window and a 12 stage box and linear-focused style dynode structure~\cite{HamPhoto}.  The photocathode, made of low temperature bialkali, is 76 mm in diameter with an effective diameter of 64 mm giving 71\% active area coverage. The dynode structure allows for good collection efficiency, and pulse linearity~\cite{PMTBook}. The dynodes and photocathode are enclosed in a kovar metal body support structure with a ceramic stem and kovar leads shown in figure~\ref{fig:r11410_picture}. 

Intended to be used in extremely low background experiments, the R11410-10 PMT is made from very low radioactive materials.  The manufacturer quoted radioactivities are: 3.3 mBq/piece for $^{238}$U, 2.3 mBq/piece for $^{232}$Th, 5.7 mBq/piece for $^{40}$K and 9.1 mBq/piece for $^{60}$Co. Further measurements with different techniques -- such as high purity germanium screening (HPGe), mass spectrometry (ICP-MS), and neutron activation (NAA) -- are ongoing.

The R11410--10 has been tested in a 6 Atm pressurized water tank and after 90 hours no structural damage or change in performance has been observed.\footnote{Future liquid xenon ton scale detectors on the order of 1 m height will have a maximum pressure of 3 atmospheres~\cite{NobleLiquid}.}

\begin{figure}[htb]
\begin{center}
\includegraphics[height = 80mm]{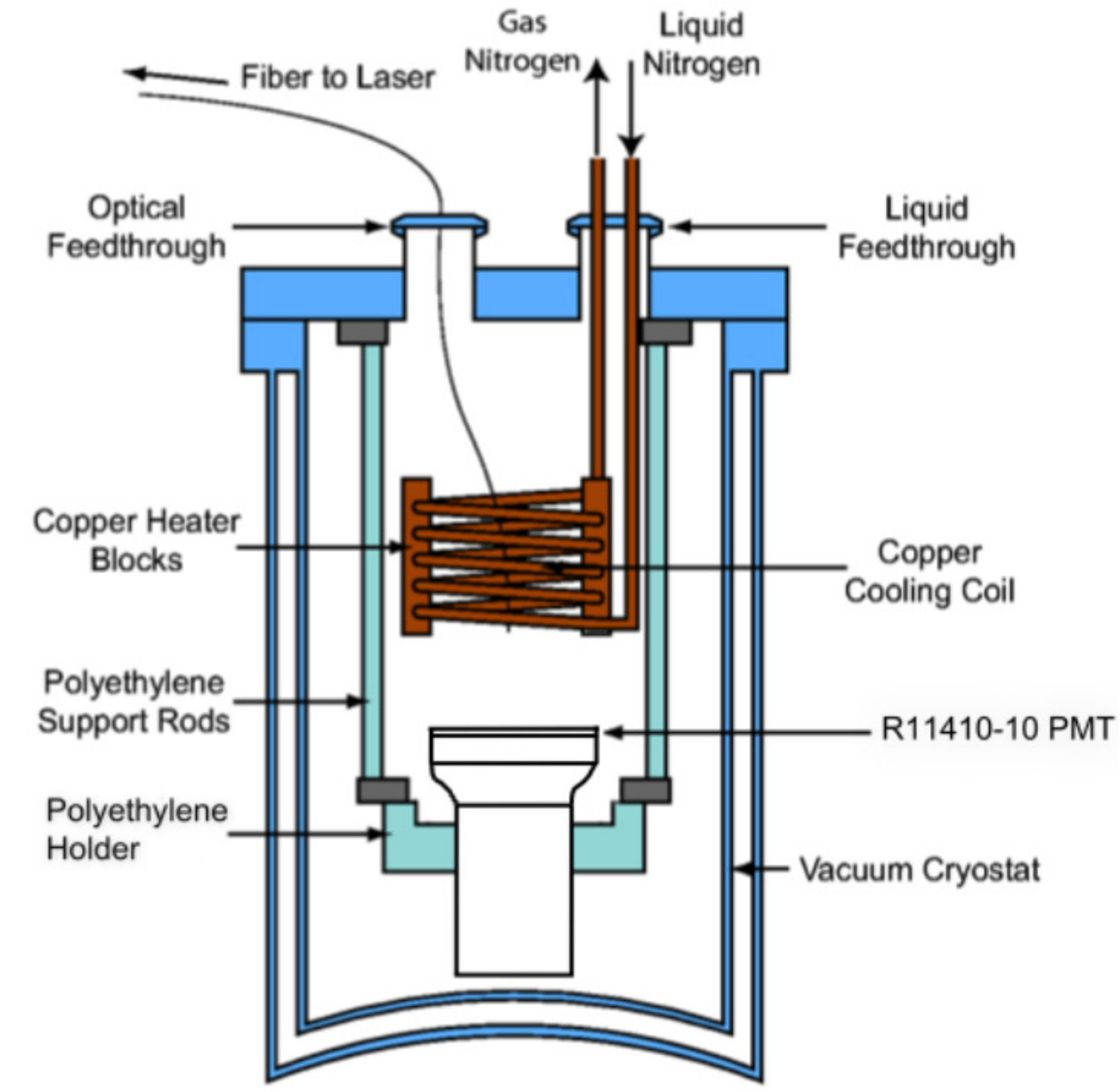}
\end{center}
\caption{
 \label{fig:setup}
 Schematic of the liquid nitrogen cooling setup for the R11410--10 PMT at UCLA.
}
\end{figure}

\section{UCLA Experimental Setup}
\label{sec:setup}
For all measurements performed at UCLA, the setup shown in figure~\ref{fig:setup} is employed. The outer infrastructure consists of a double-walled stainless steel cryostat designed to be both light and vacuum tight. For low temperature operation, a vacuum is pumped in the cryostat using an oil free pumping station\footnote{Pfieffer Vacuum Hi-Cube Eco-3 Pumping Station with a diaphragm pump and a turbomolecular pump.} and then filled with dry gas nitrogen. The gas is cooled by convection through the constant flow of liquid nitrogen in the copper coil regulated by a mass flow controller. Heaters attached to the coil are connected to a Proportional-Integral-Derivative (PID) controller\footnote{Omega Model CN8201 Temperature Controller.}, which uses feedback from Resistance Temperature Detectors (RTDs) placed in contact with the PMT to control the temperature. This system is able to maintain a temperature down to -150 $^{\circ}$C with $0.1 ^{\circ}$C accuracy.

\begin{figure}[htb]
\begin{center}
\includegraphics[height = 28mm]{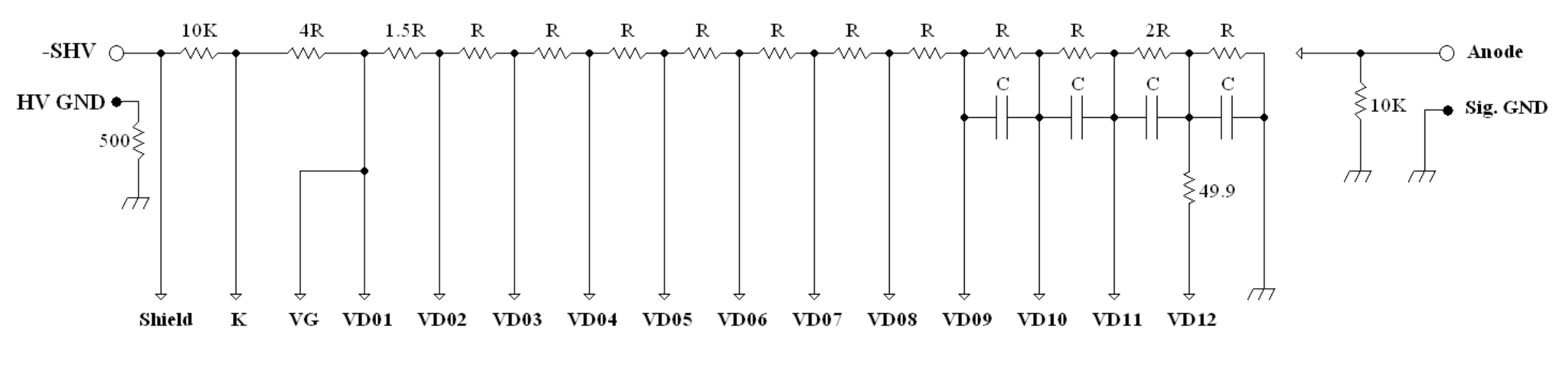}
\end{center}
\caption{
 \label{fig:pmt_base}
 Base schematic for the R11410--10 PMT used in all measurements which require standard operation. In this diagram, $\mathrm{R}=1\mathrm{M}\Omega$ and $\mathrm{C}=4.7 \mathrm{n}F$ rated at 250 V.
}
\end{figure}

Several BNC and SHV-10 feedthroughs are connected for signal readout and high voltage application. For the majority of tests performed except where noted, the standard negative base was used (figure~\ref{fig:pmt_base}) with a typical operating voltage\footnote{Stanford Research Systems Model PS325 2500V-25W.} of -1750 V. A fiber is inserted in the setup, which allows for an external laser\footnote{Hamamatsu Laser Head M10306-30 402 nm 53 ps pulse width 269 mW.} and pulser\footnote{Hamamatsu Picosecond Light Pulser C10196.} to be used for measurements and calibrations. Data from pulsed measurements is recorded using an oscilloscope\footnote{LeCroy Oscilloscope WaveRunner 204Mxi-A 2GHz, 10 GS/s.} and current measurements are performed with a picoammeter\footnote{Keithley Picoammeter 6485.}.

\section{Signal Response}
\label{sec:signal}

\begin{figure}[htb]
\begin{center}
\includegraphics[height=75mm]{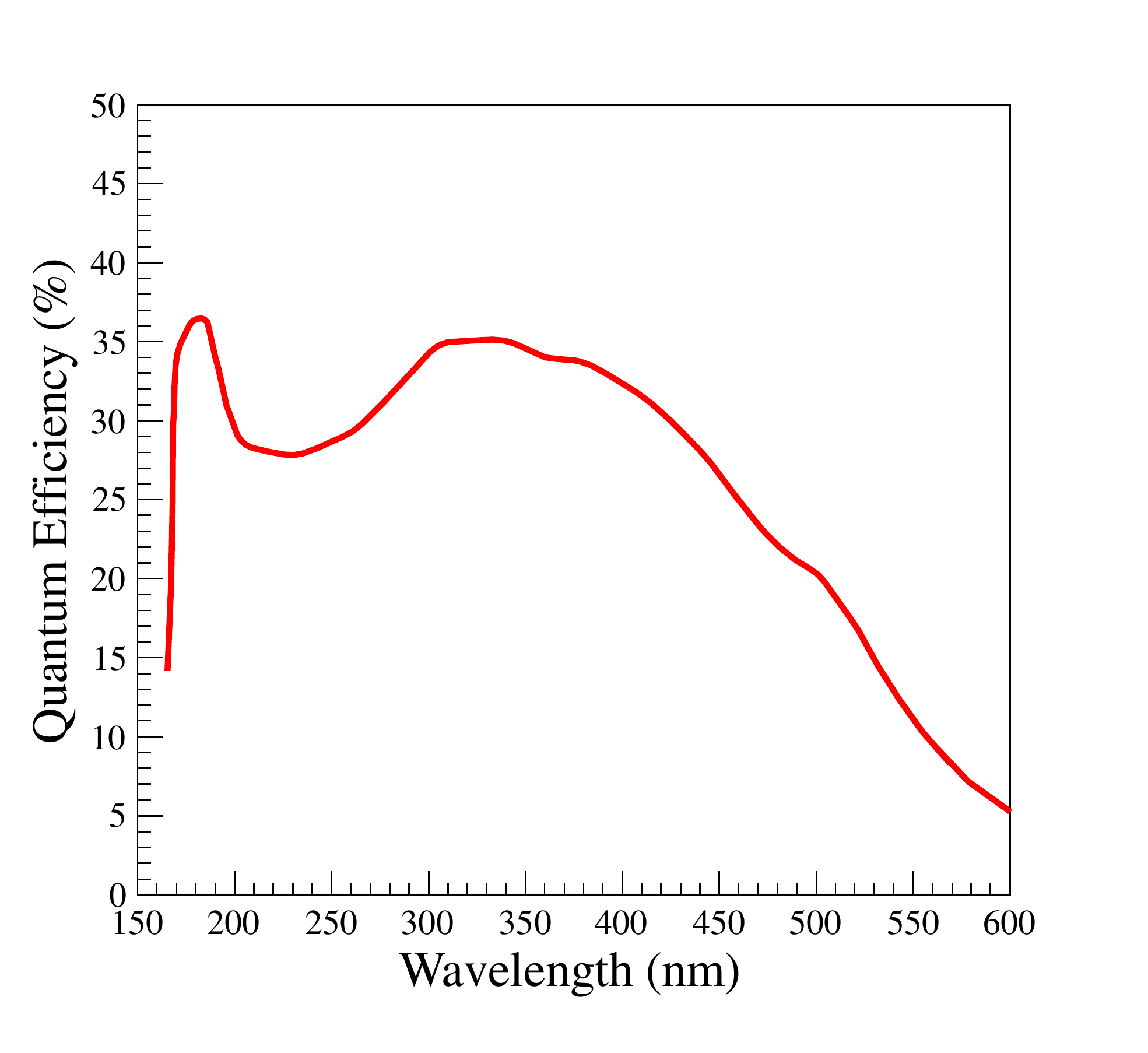}
\end{center}
\caption{
 \label{fig:QEfig}
 Quantum efficiency versus wavelength for a typical R11410-10 model PMT. The maximum quantum efficiency of 35\% occurs near 178 nm, the scintillation wavelength of xenon.  Below 170 nm, there is a sharp cutoff due to the opacity of the quartz window.
}
\end{figure}

\subsection{Quantum Efficiency}
To maximize the detector light yield (number of photoelectrons per unit energy deposited) and effectively lower the energy threshold, a high quantum efficiency (QE) is necessary. 
Hamamatsu has measured the QE at room temperature by comparing the response of the R11410-10 PMT to that of a standard PMT, which was calibrated with a NIST standard UV sensitive photodiode. Figure~\ref{fig:QEfig} shows a typical QE dependence on wavelength with a maximum of about 35\% around 178 nm and 30\% between 300 and 400 nm. Among production line PMTs, these maxima vary by about 3\%. The high quantum efficiency around 400 nm, the shifted wavelength for argon scintillation light, allows this same device to be applicable for liquid argon detectors.

\subsection{Single Photoelectron Gain}
\label{sec:SPE}
An inherent detector threshold on a signal is required to maintain a satisfactory signal to noise ratio. A high gain increases the S/N ratio by allowing for more single photoelectron pulses to pass the threshold. 

\begin{figure}[htb]
\begin{center}
\includegraphics[height=75mm]{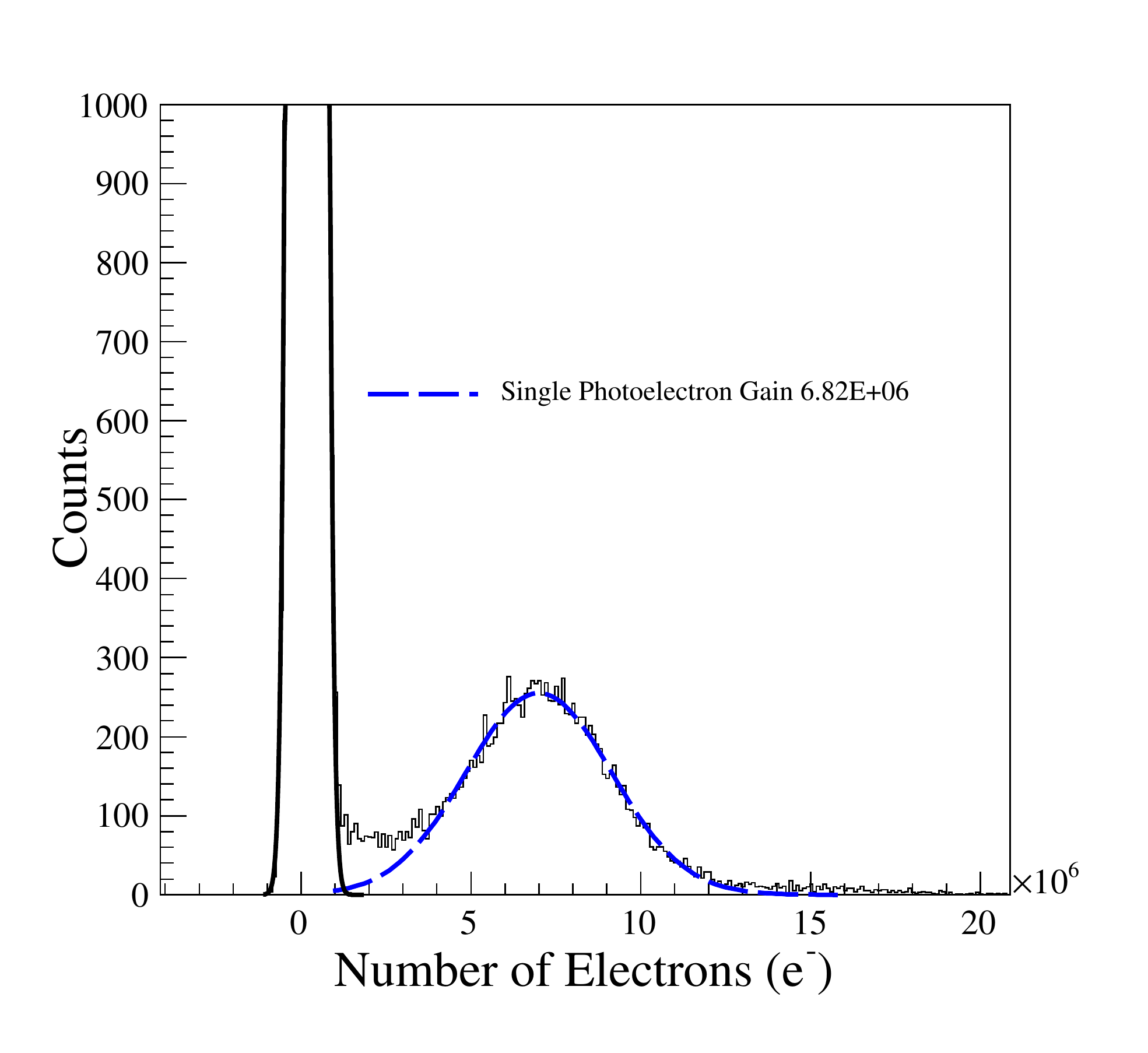}
\end{center}
\caption{ 
 \label{fig:SPE}
 Single photoelectron gain spectrum at room temperature and -1750 V of the R11410-10 PMT. The pedestal and single photoelectron peaks are fitted with gaussians resulting in a single photoelectron gain of $6.82\times 10^{6} $. 
}
\end{figure}

To measure the single photoelectron gain, the PMT is placed in the setup shown in figure~\ref{fig:setup} at the desired temperature. For the results shown in figure~\ref{fig:SPE}, the PMT with -1750 V applied is held at room temperature in a gas nitrogen environment.

Light is introduced to the cryostat and PMT through the fiber by means of the pulsed picosecond laser, which also triggers the oscilloscope. In order to measure the single photoelectron gain, the light level is adjusted such that only one out of every ten laser pulses results in the observation of a  photoelectron signal from the PMT. Since this physical process will be distributed according to a Poisson distribution, of the 10\% of signals which contain photoelectron pulses only 1 out of 20 $\left(\frac{Pois(\mu=0.1,x=1)}{Pois(\mu=0.1,x>1)}\right)$ will have more than 1 photoelectron. As a result, a relatively unpolluted sample of single photoelectrons is obtained. 

A time window after the triggered laser pulse is defined where the area of the signal pulse is measured. The typical window size used is on the order of several single photoelectron pulse widths (section~\ref{sec:timing}), about 50 ns. The area of signal pulse is converted into a charge in units of electrons. The absolute gain is then obtained by the background subtracted single PE distribution.


The results for a typical PMT are shown in figure~\ref{fig:SPE}. The value of the absolute gain at -1750 V is $6.82\times 10^{6}$ with a 5\% systematic error, which provides gain sufficient to measure single photoelectrons above noise. The gain was also measured at liquid xenon temperature and found to be within the systematic error of the room temperature measurement and stable over several days of operation. 



\subsection{Timing Properties}
\label{sec:timing}

\begin{figure*}[htb]
\begin{center}
\includegraphics[height=60mm]{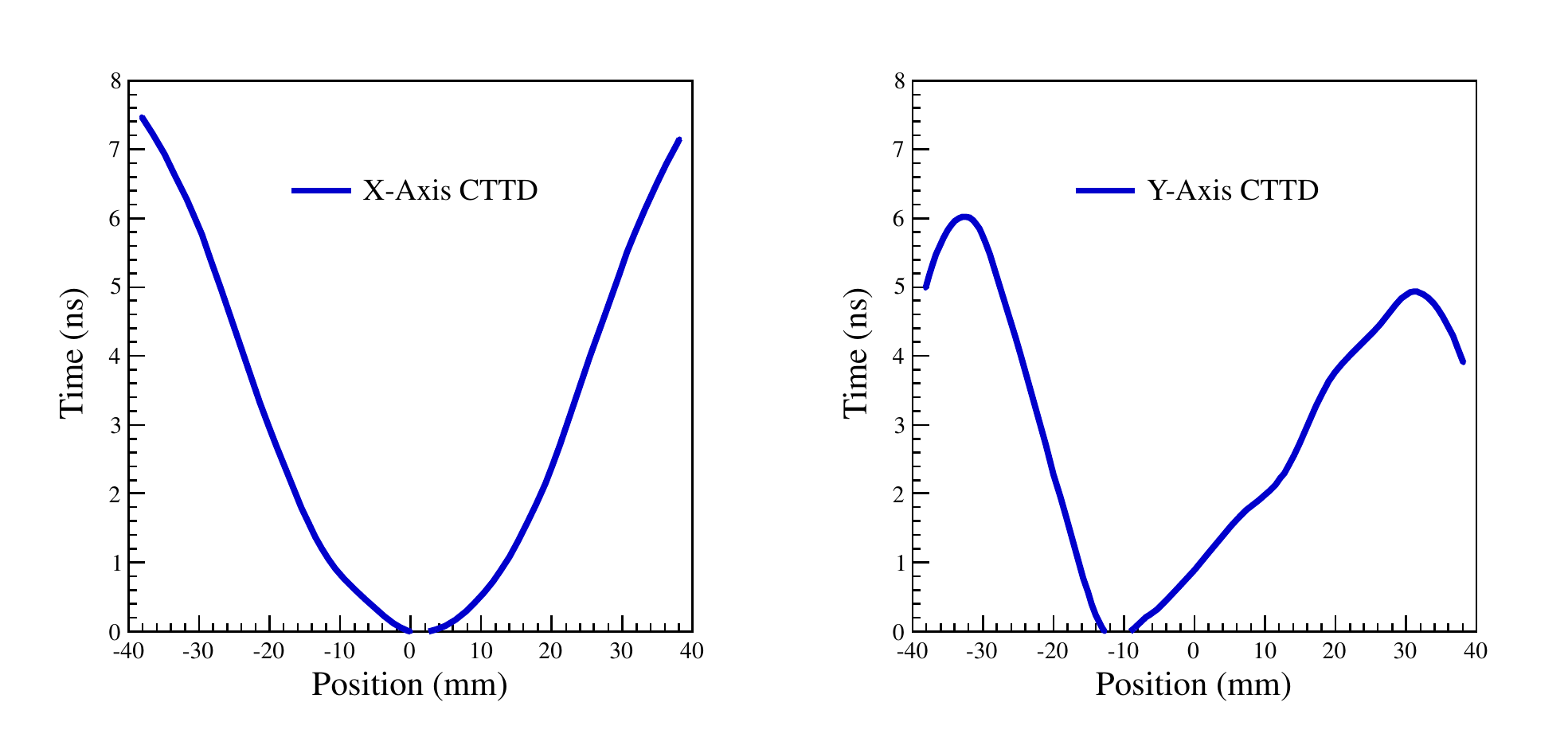}
\end{center}
\caption{
 \label{fig:CTTD}
 Cathode transit time difference for positions along the X and Y axes of the R11410-10. The largest deviations occur near the edges where small variations in the internal structure cause the electron travel time to increase. The minima is set to zero, as this is a relative measurement.
}
\end{figure*}



The single photoelectron pulse timing characteristics are measured at UCLA in the same setup as in section~\ref{sec:SPE} with the PMT operating at -1750 V. The measurements resulted in a rise time of $2.6\pm 0.4$ ns (10\% to 90\%), a fall time of $9.7\pm 0.5$ ns (90\% to 10\%) and a pulse width of $7.2\pm 0.6$ ns (FWHM). The transit time spread (TTS), using a diffuser to uniformly illuminate the photocathode, is determined by the FWHM of the distribution of times between the trigger and the maximum of the single photoelectron pulse. At -1750 V, the TTS \footnote{The TTS is dependent upon the voltage, which affects the electron travel time} was found to be $6.0\pm0.3$ ns. 

\begin{figure*}[htb]
\begin{center}
\includegraphics[height=60mm]{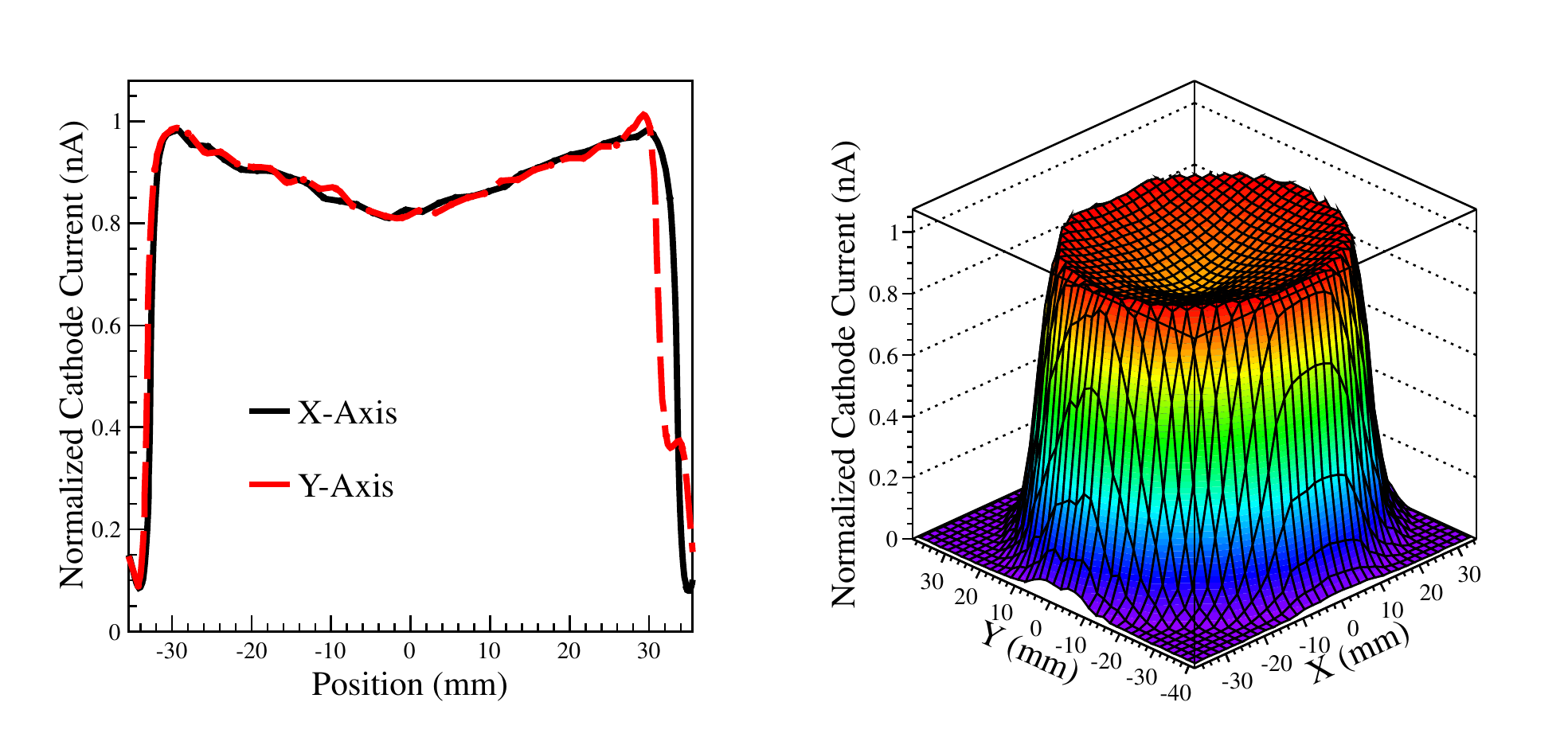}
\end{center}
\caption{
 \label{fig:uniformity}
 Position dependent relative uniformity for the photocathode of the R11410-10 PMT. The PMT output signal is uniform to within 20\% across the whole quartz window.
}
\end{figure*}

\begin{figure*}[htb]
\begin{center}
\includegraphics[height=60mm]{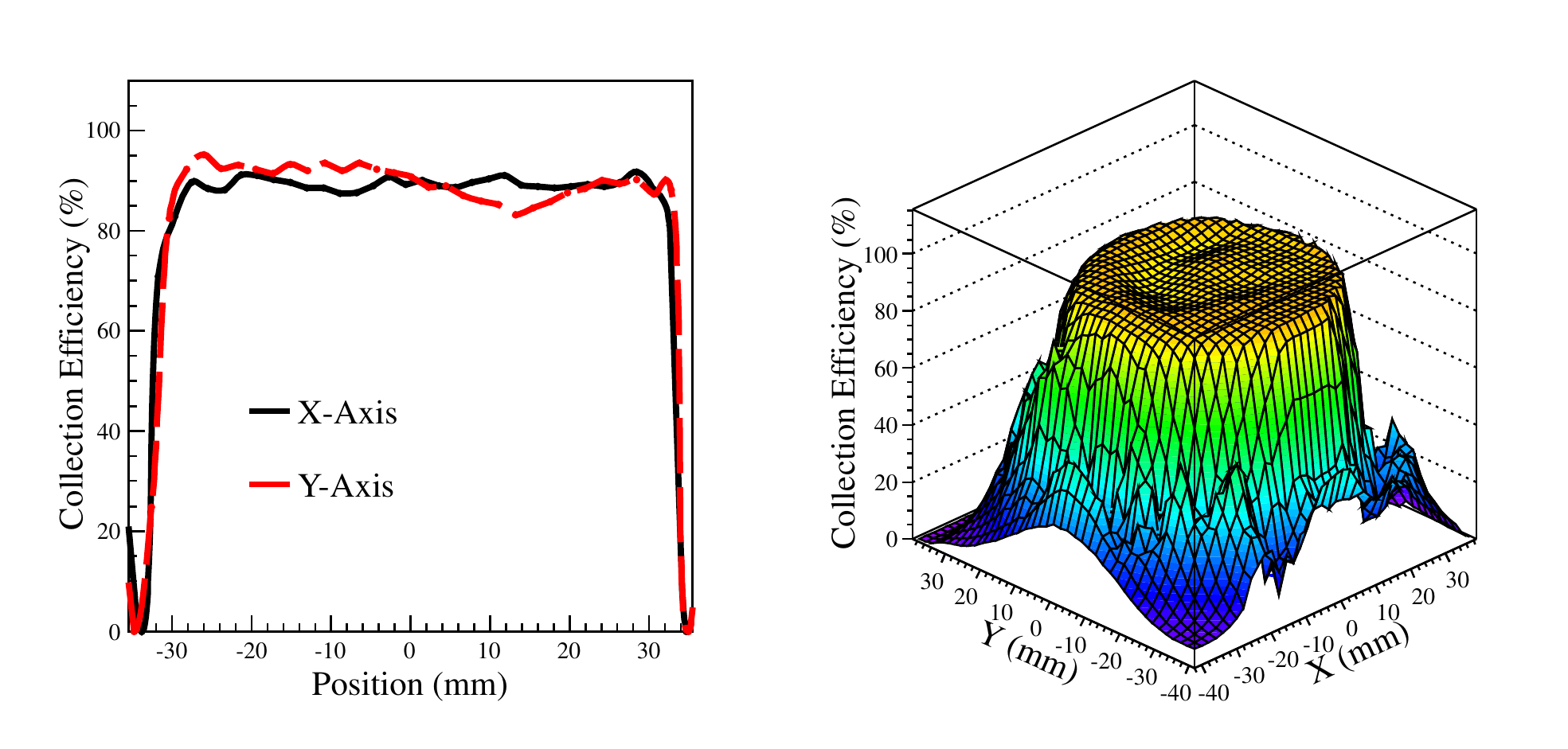}
\end{center}
\caption{
 \label{fig:CE}
 Position dependent relative collection efficiency R11410-10 PMT. The collection efficiency is uniform to within 10\% over the surface.
}
\end{figure*}

The cathode transit time difference is measured using a similar method to transit time spread. The input laser light is focused at designated points along the X and Y axes (with respect to the first dynode) and measuring the mean of the distribution of times between the trigger and maximum of the single photoelectron pulse. The results are shown in figure~\ref{fig:CTTD}. Near the edges, deviations greater than 5 ns may occur as the transit time is highly dependent upon incoming position. The symmetry along the X axis is expected as this cuts across the first dynode. However, the Y axis scan shows a skewed feature, which is evidence of the sloped dynode structure.


\subsection{Uniformity}

To quantify the uniformity of the PMT surface, an LED of 402 nm is focused with a 1 mm spot size on the quartz window. A two axis scanner coplanar to the photocathode is used. The current, either photocathode or anode, is read out by the picoammeter and then normalized by the largest current, which occurs near the edges for the R11410--10  (figure~\ref{fig:uniformity}). The photocathode and anode are both uniform to 20\% across the whole surface. In comparison to the position dependence of the transit time, this signal magnitude uncertainty is much lower. 

Taking the ratio between the anode and photocathode uniformity scans and accounting for the gain of the PMT results in a measure of the collection efficiency. The result in figure~\ref{fig:CE} demonstrates that the collection efficiency is greater than 80\% for the entire surface and uniform to within 10\%.

\subsection{Cathode Linearity}
\label{sec:cathodelinearity}
The cathode linearity quantifies the range over which the photocathode PMT response to input light is linear. Though not vital for the low energies expected for dark matter signals, a high linearity allows for additional detector calibration methods through the use of higher energy sources. 

At Hamamatsu, the cathode linearity setup includes: a tungsten lamp coupled with a diffuser to uniformly illuminate the photocathode of the R11410-10 device; a set of four shutters and a neutral density filter wheel to vary the input light intensity from the lamp; and a picoammeter to measure the current from the PMT. The photocathode of the PMT is held at  -300 V, while the dynodes and anode are attached to the readout at ground. Electrons that are emitted from the photocathode are collected and readout by the picoammeter. 

For a given temperature and uniform light intensity, each of the four shutters is opened individually with the others closed while the resulting photocathode current is recorded from the picoammeter. The total signal with each of the four shutters open is also recorded. The light intensity is then increased or decreased using the neutral density filter and the shutter process is repeated. 

Using the recorded data from each of the shutters and the total signal data, for each light intensity, a deviation from linearity is determined:

$$\Delta Linearity = \frac{I_{total} - \Sigma^{4}_{n=1}I_{n}}{\Sigma^{4}_{n=1}I_{n}}.$$

\begin{figure}[htb]
\begin{center}
\includegraphics[height=75mm]{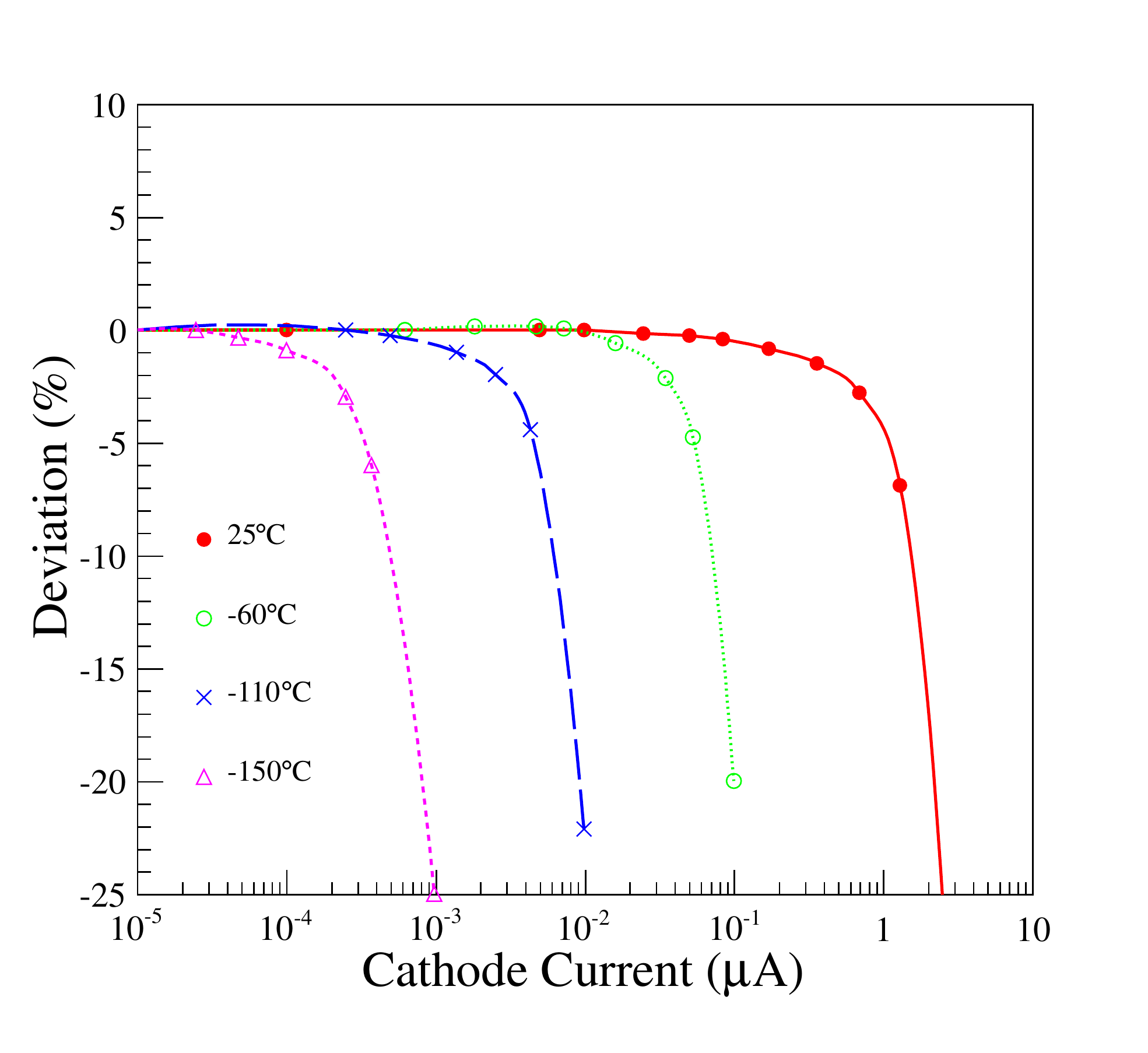}
\end{center}
\caption{ 
 \label{fig:CathodeLinearity}
 Cathode linearity of the R11410-10 PMT versus output photocathode current and temperature. The cathode becomes nonlinear at lower temperatures at lower output currents. The cathode is linear to within 5\% at liquid xenon temperatures up to 2 nA.
}
\end{figure}

The results of this measurement are shown in figure~\ref{fig:CathodeLinearity}. The cathode linearity shows a strong dependence upon the temperature of the environment, where lower temperatures cause the cathode to become nonlinear at lower currents due to the increasing resistivity of the bialkali metal~\cite{PMTBook}. For operation in liquid xenon near $-110 ^{\circ}$C, the cathode is linear (within 5\%) up to 2 nA.

\begin{figure}[htb]
\begin{center}
\includegraphics[height=75mm]{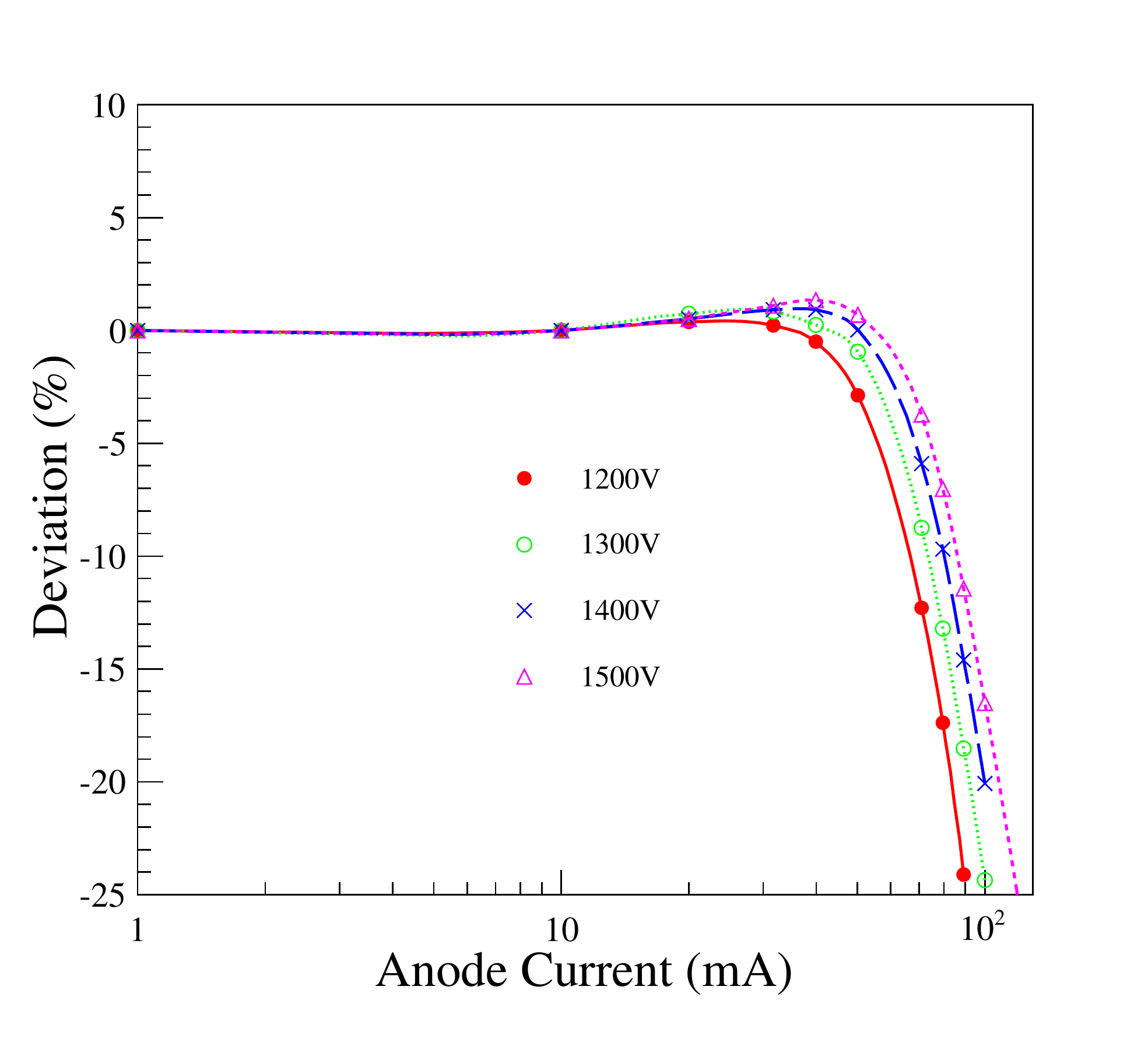}
\end{center}
\caption{
 \label{fig:AnodeLinearity}
 Percentage deviation from perfect linearity versus the output anode current (mA). The R11410-10 device is linear to within 5\% up to about 80 nA.
}
\end{figure}

\subsection{Anode Linearity}
\label{sec:AnodeLinearity}
The loss of linearity in a photomultiplier tube is due to saturation of the photocathode, as detailed in Section~\ref{sec:cathodelinearity}, and the space charge effect between the last dynode stage and the anode. The interplay between the photocathode and anode linearity is determined by the operation gain of the photomultiplier tube and the temperature of the environment. 



The anode linearity setup is the same as the cathode linearity, except that the former is a pulsed measurement, using the normal readout through all the dynode stages. For the R11410-10 PMT, the anode linearity has been measured for several different voltages as shown in figure~\ref{fig:AnodeLinearity},  and is linear within 5\% to above 80 mA anode current for all voltages.  These results with respect to the cathode linearity measurements show that nonlinearity in the photocathode dominates at low temperatures, while nonlinearity in the anode at high temperatures becomes prominent.


\section{Noise characterization}
\label{sec:noise}

\subsection{Dark Current and Gain}
Application of high voltage to the dynodes of the PMT results in both an increase in gain and dark current. This dark current can stem from a number of features including field emission, leakage current, and thermionic emission~\cite{DarkCurrent}. The gain and dark current have been measured to quantify their dependence upon the high voltage applied to the R11410--10. 

\begin{figure}[htb]
\begin{center}
\includegraphics[height = 75mm]{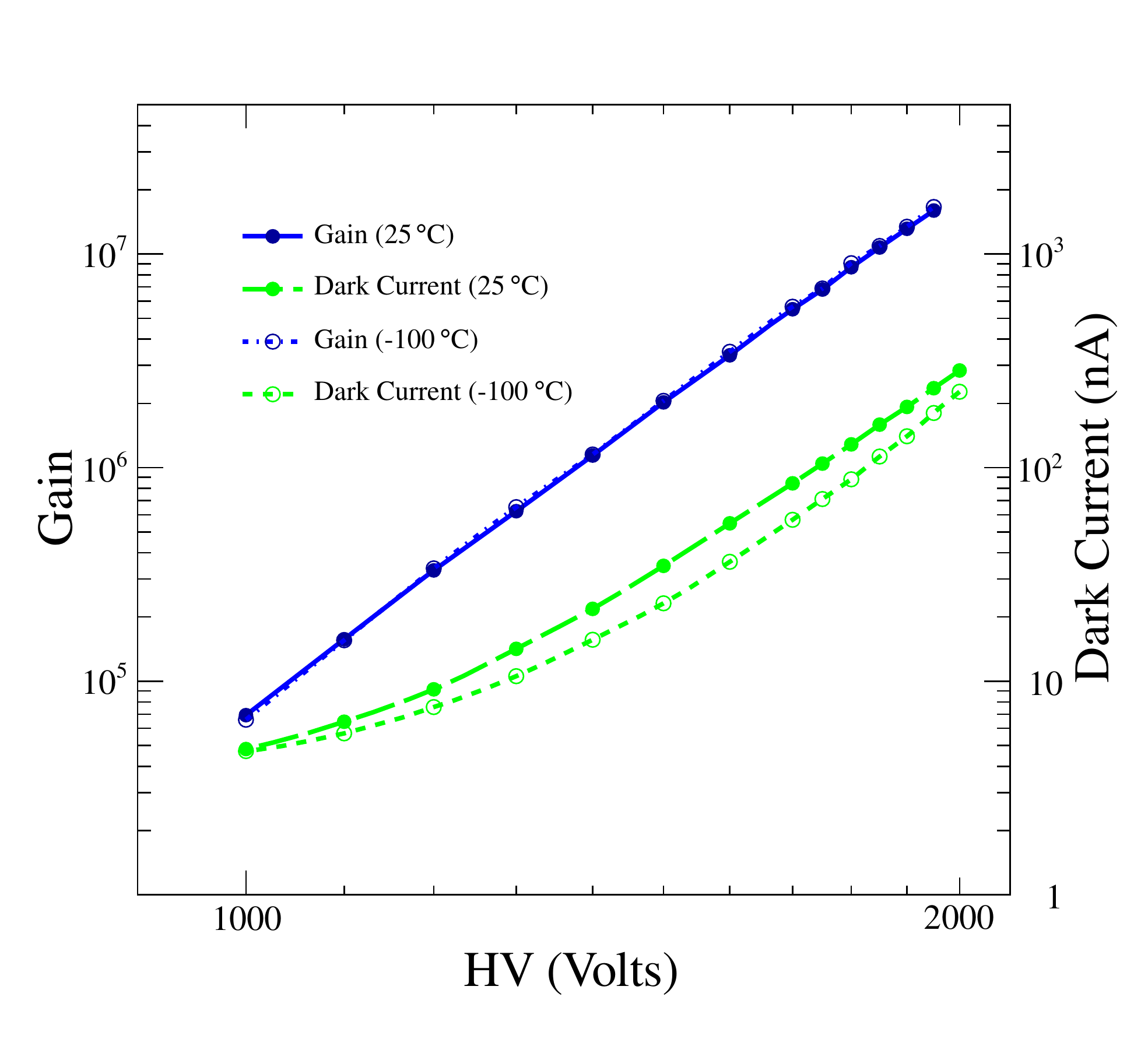}
\end{center}
\caption{
 \label{fig:darkcurrent}
 Dark current (green) and gain (blue) versus high voltage of the R11410-10 PMT for both room temperature and liquid xenon temperature. Two features of the dark current curve are readily seen at low voltages (-1000 to -1200 V) and high voltages (-1400 to -1900 V). The gain exhibits a power law as expected from the dynode multiplication. The dark current curve at low temperature has been normalized to the room temperature value through the addition of an offset.
}
\end{figure}

To measure the dark current, the PMT is placed in the cryostat for several hours before operation to reduce any effects the ambient light may have. For each high voltage applied between -900 and -1900 V, the anode current from the picoammeter is read in 100 V steps. 

The gain dependence on the high voltage applied is measured by pulsing the laser head with a light intensity corresponding to hundreds of photoelectrons, but below the nonlinear region of the phototube. The relative gain is recorded by measuring the average area of 10000 waveforms at each voltage applied. The absolute gain is given by the single photoelectron gain at -1750 V (section~\ref{sec:SPE}) and scaled for each voltage based on the relative measurement.

The results of the dark current and gain measurements are shown in figure~\ref{fig:darkcurrent}. The gain curve dependence on the high voltage follows a power law as expected due to the dynode multiplication stages. The dark current shows several trends with the voltage dependence. The trend at lower voltages is attributable to the presence of a leakage current through the device. As the voltage increases, a second component appears, which is due to the constant thermionic emission from the photocathode~\cite{PMTBook}. Above this range, field emission and scintillation in the glass begin to dominate. The operable range which gives the best signal to noise ratio of the phototube is defined by the region where the slopes of the dark current and gain curves are the same, in this case taken to be -1750 V, where mostly thermionic emission is prominent. 


\begin{figure}[htb]
\begin{center}
\includegraphics[height=75mm]{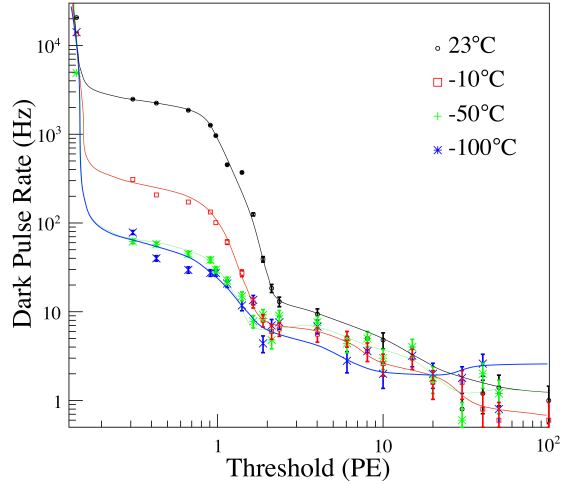}
\end{center}
\caption{
 \label{fig:darkcount}
 Temperature variation of the dark pulse rate versus threshold in photoelectrons. The rate at 1 PE threshold drops steeply with temperature until 50 $^{\circ}$C to approximately 50 Hz. Additionally, the dark pulse rate above 2 PE threshold is less than 10 Hz. Statistical error bars are shown. 
}
\end{figure}

\subsection{Dark Pulse Rate}
Dark pulses result from the spontaneous thermionic emission of single electrons from the photocathode, intrinsic radioactive impurities in the materials and cosmic ray signals. In a detector, a coincidence requirement between PMTs is imposed to reduce effects of single dark pulses. The remaining background from dark pulses is due to an accidental rate between multiple dark pulses and/or a true scintillation signal. This background can be understood through the absolute dark pulse rates measured here.


To measure the dark pulse rate, a random trigger window on the oscilloscope is opened for 10 microseconds, during which the number of pulses that pass a designated threshold are counted. Several thousand triggers are accumulated to minimize the effects of statistical fluctuations. The dark pulse rate is recorded for several thresholds and temperatures ranging from $25 ^{\circ}$C to $-100 ^{\circ}$C. 

The results of the dark pulse measurements at -1750 V are shown in figure~\ref{fig:darkcount}. The threshold has been converted from volts to photoelectrons using a pulse height distribution for single photoelectron signals at each temperature. At thresholds below 2 PE, thermionic emission accounts for the majority of dark counts, which results in the large temperature dependence of the rate; the dark pulse rate in the region around one photoelectron threshold decreases with temperature from about 1 kHz until fluctuations in the leakage current begin to dominate near $-50 ^{\circ}$C with a rate of 50 Hz. 

Above this threshold, the rate is observed to decrease significantly and becomes temperature independent. Thermionic emission becomes negligible as cosmic rays and material impurities make up the majority of the 10 Hz rate. 

\begin{figure}[htb]
\begin{center}
\includegraphics[height=75mm]{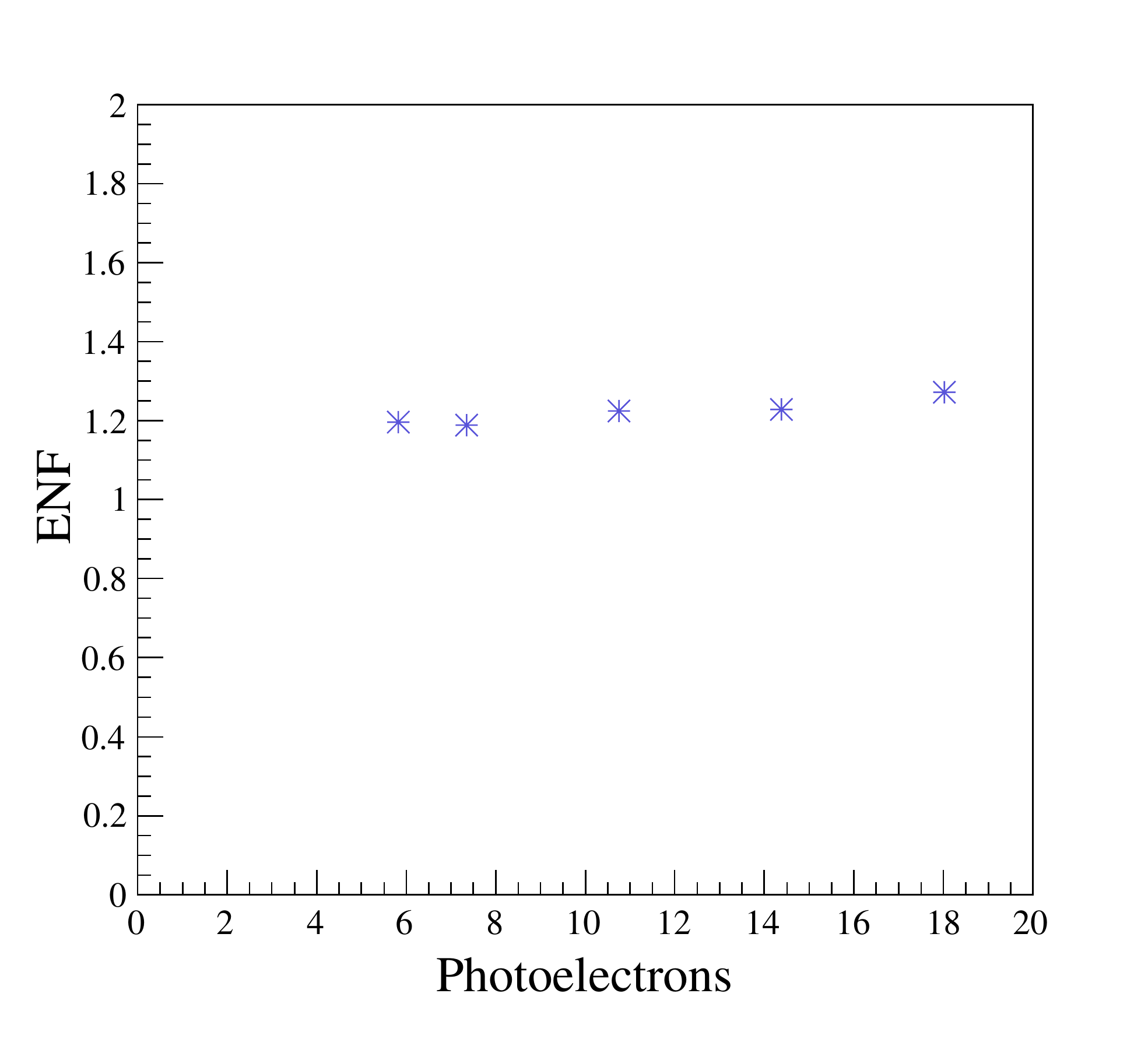}
\end{center}
\caption{
 \label{fig:ENF}
 Excess noise factor of the R11410--10 PMT versus the signal input at $-100 ^{\circ}$C. The ENF is stable around 1.2 for signals smaller than 20 photoelectrons.
}
\end{figure}

\subsection{Excess Noise Factor}
The excess noise factor (ENF) is a measure of the deviation of the observed photodetector resolution to the expected resolution based on Poisson statistics. It is caused primarily by noise added during each step of the multiplication process in the dynode stages~\cite{ENF}. 






To measure the excess noise factor, the laser is pulsed at the most stable operating conditions to minimize the fluctuations of the light input. Using a set of inline filters\footnote{Thorlabs NE500A Series Mounted Absorptive Neutron Density Filters.}, the light intensity is varied at several values below 20 PE, where the excess noise factor dominates over variations in light intensity caused by the setup. For each input light intensity, the area for many pulses in units of photoelectrons is considered. The excess noise factor is estimated as follows:

$$ENF = \frac{\sigma^{2}_{obs}}{\sigma^{2}_{exp}} = \frac{\sigma^{2}_{obs}}{\left(\sqrt{N_{PE}}\right)^{2}}$$

\noindent where $\sigma_{obs}$ is the standard deviation of the distribution and $N_{PE}$ is the mean of the distribution. 

The resulting excess noise factor for photoelectron values less than 20 is shown in figure~\ref{fig:ENF} for an operation voltage of -1750 V at liquid xenon temperature. A measured ENF of 1.2 results in an energy resolution of approximately 10\% higher than expected due to only Poisson fluctuations. 

\begin{figure}[htb]
\begin{center}
\includegraphics[height=75mm]{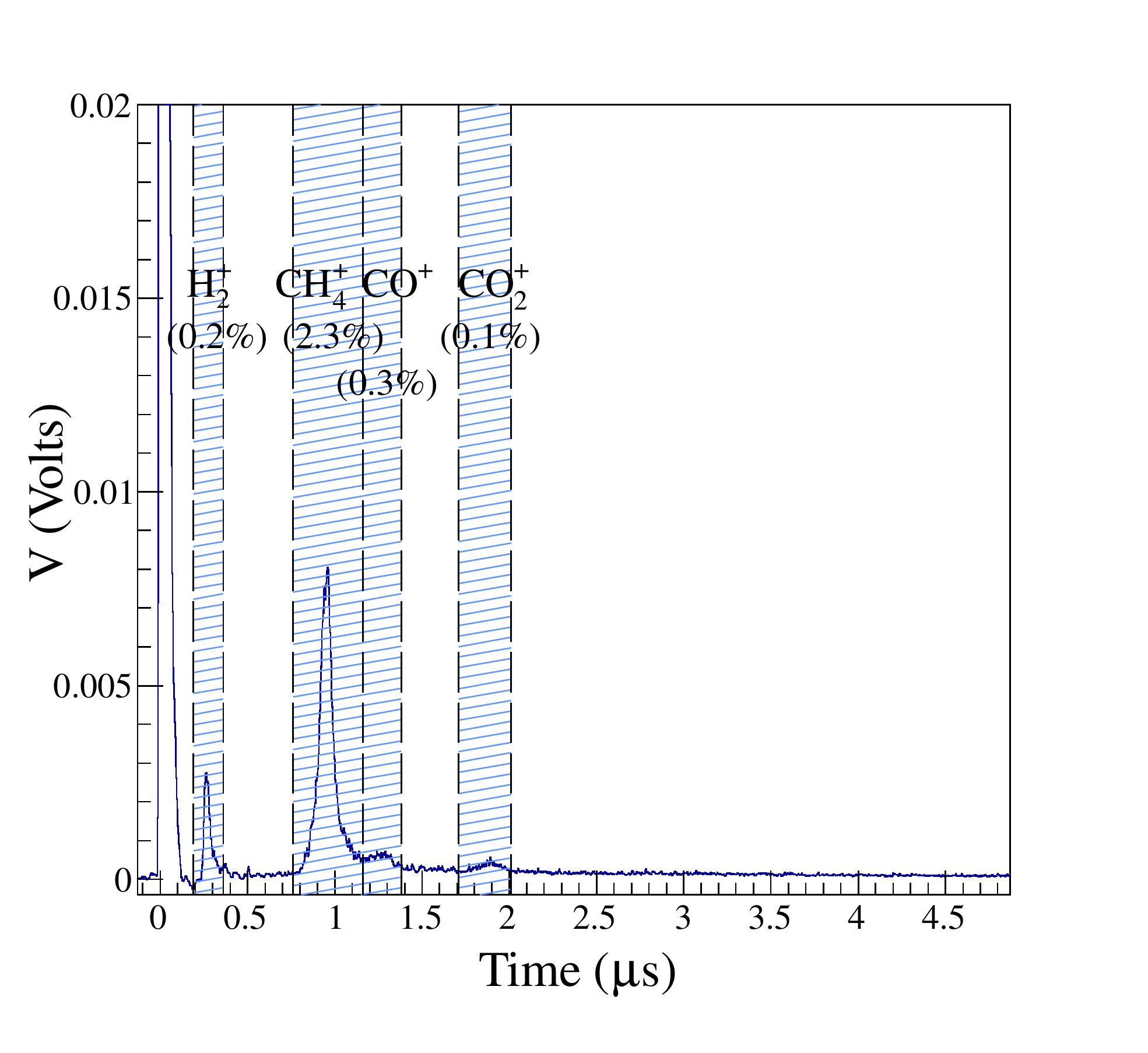}
\end{center}
\caption{
 \label{fig:afterpulsing}
 Averaged timeline of 10000 waveforms after a 500 PE main pulse. Four main afterpulse regions are visible, which contribute to an overall afterpulse rate of 4.9\%. 
}
\end{figure}

\subsection{Afterpulsing}
Afterpulsing is the phenomena by which electrons emitted from the photocathode ionize impurities in the vacuum and dynode structure, which then drift back to the photocathode and produce a secondary signal on the timescale of microseconds~\cite{Afterpulse}. This results in the loss of charge from the initial signal while possibly producing accidental coincidences of secondary pulses with a true energy deposit in the detector volume. 

The afterpulsing is measured by illuminating the PMT at -1750 V with a large light signal of approximately 500 PE (300 pC). By averaging 10000 waveforms on the oscilloscope as shown in figure~\ref{fig:afterpulsing}, regions after the main signal are visible above the baseline which are attributed to various elements. The characteristic afterpulse time is dependent upon the individual element and its origin: 

$$A.T. = \sqrt{\frac{2md}{qV}}$$

\noindent where the afterpulse time is related to $m$, the mass of the ion, $d$, the mean distance between the cathode and the first dynode (where most of the impurities are expected to reside), and $V$, the voltage between the cathode and first dynode. We infer the four afterpulse regions to result from ions of hydrogen ($H_{2}^{+}$), methane ($CH_{4}^{+}$), carbon monoxide ($CO^{+}$), and carbon dioxide ($CO_{2}^{+}$).

The total afterpulse rate is defined as:

$$ APR = \frac{\Sigma Q_{AP}}{Q_{MP}}$$

\noindent where $Q_{AP}$ is the amount of charge in the total afterpulse region (after 200 ns), and $Q_{MP}$ is the charge in the primary pulse. From the averaged waveform, the proportion of charge in the afterpulse region to the primary signal is found to be 4.9\%, a majority of which is due to the peaks from hydrogen (2.3\%), methane (0.3\%), carbon monoxide (0.2\%), and carbon dioxide (0.1\%). At -100$^{\circ}$C, this effect is reduced to 1.8\% likely caused by impurities adhering to inner surfaces of the PMT as the temperature decreases. For smaller signals, the absolute rate of afterpulsing becomes negligible due to the quantization of charge afterpulses.

\begin{figure}[htb]
\begin{center}
\includegraphics[height = 40mm]{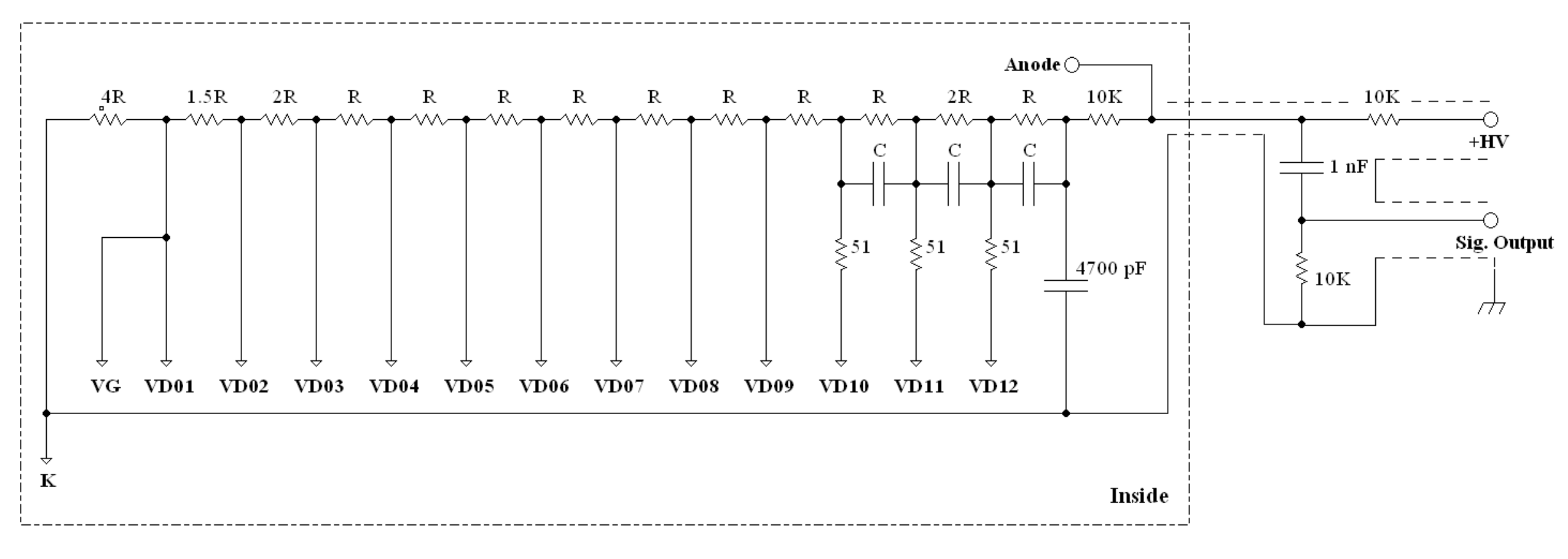}
\end{center}
\begin{multicols}{2}
\begin{center}
\includegraphics[height = 40mm]{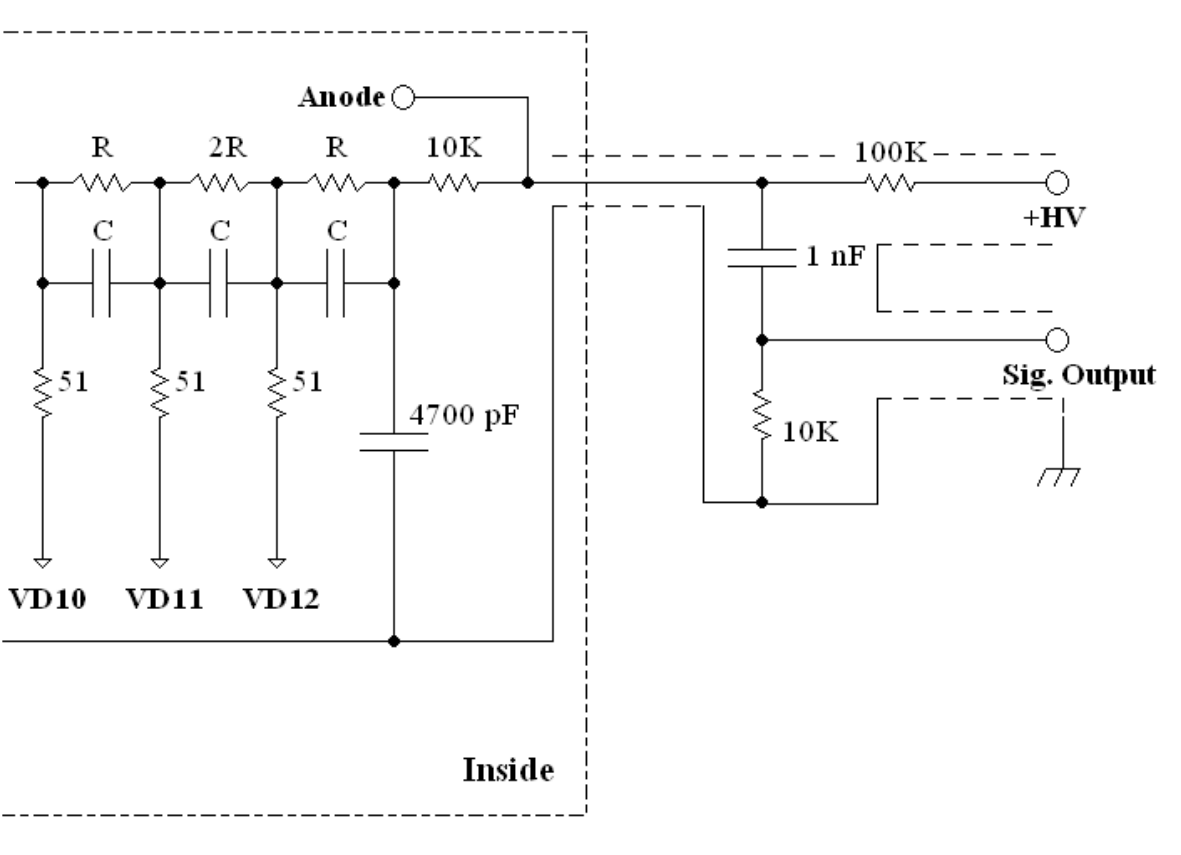}
\end{center}

\begin{center}
\includegraphics[height = 40mm]{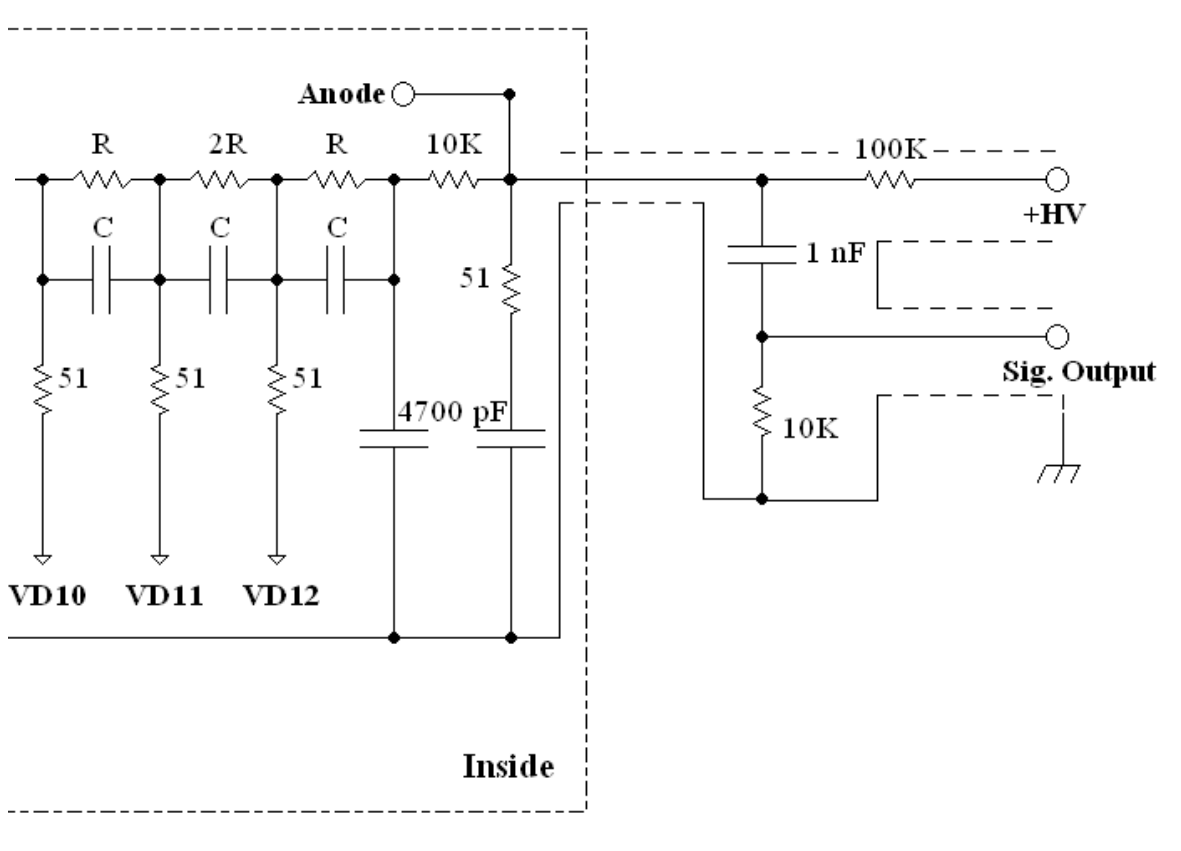}
\end{center}
\end{multicols}
\caption{
\label{fig:bases_positive}
  Three base schematics used for single cable operation with varying decoupling configurations. The \textit{top panel} shows the schematic with a decoupling circuit using a resistance of 10 k$\Omega$ and 1 nF capacitance. The \textit{bottom left panel} shows a larger decoupling resistor while the \textit{bottom right} configuration has an additional 51 $\Omega$ resistor and 4700 pF capacitor at the anode. 
}
\end{figure}

\subsection{PMT Base Studies}

\begin{figure}[htb]
\begin{center}
\includegraphics[height=75mm]{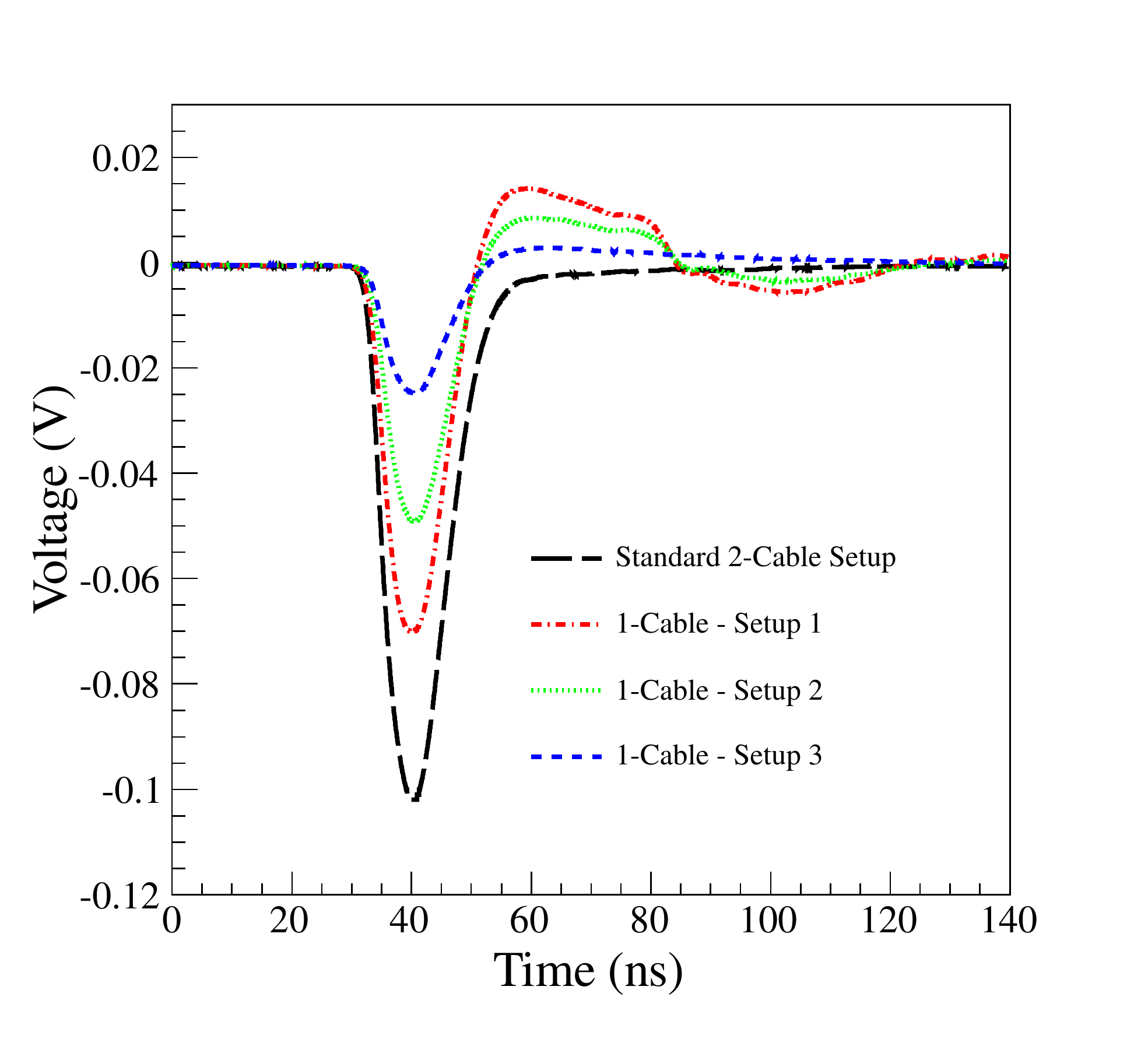}
\end{center}
\caption{
 \label{fig:waveforms}
 Sample waveforms for different cable setups. The standard setup is shown in black with alternative setups (dashed lines) showing some degradation in either signal amplitude or reflection.
}
\end{figure}

\begin{figure}[htb]
\begin{center}
\includegraphics[height=75mm]{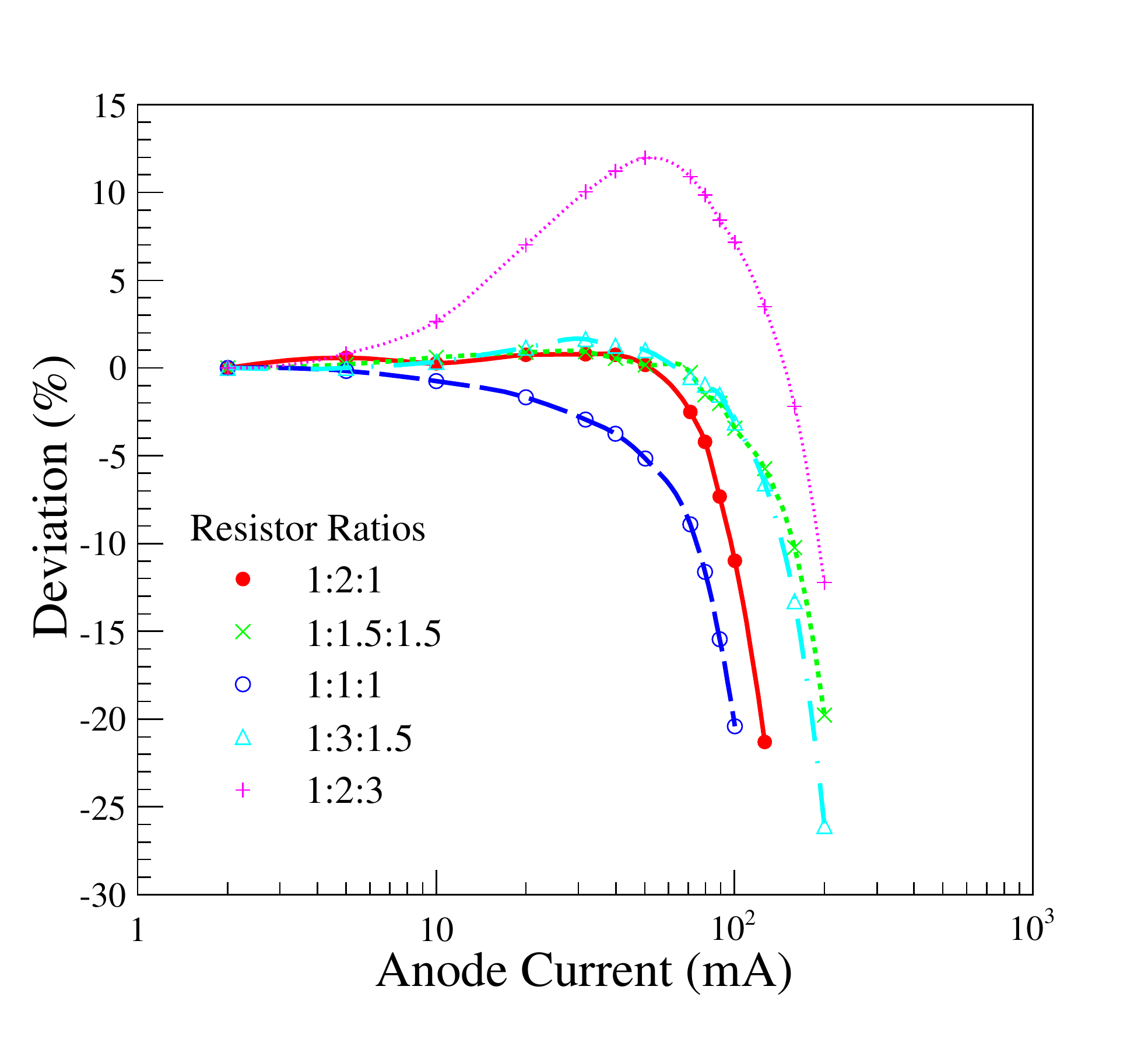}
\end{center}
\caption{
 \label{fig:HamamatsuDynodes}
 Pulsed anode linearity for the R11410-10 for various resistor ratios for the last three dynode stages. The nominal anode linearity as quoted in section~\ref{sec:AnodeLinearity} is up to 80 mA while the most optimistic scenario here achieves anode linearity within 5\% to above 100 mA.
}
\end{figure}

The development of a single cable readout system for the R11410-10 PMT is well motivated by the reduction in cabling required for a future detector as well as the change from holding the photocathode and housing at high voltage to ground simplifying support structure design. Moving from separate high voltage and signal cables to a single cable results in the need for a decoupling circuit, which introduces signal reflections at the interface. At Hamamatsu, several configurations of the base have been tested with different capacitor orientations and resistor values in order to reduce reflections and maintain the highest signal quality possible (figure~\ref{fig:bases_positive}). 

A comparison of the resulting waveforms after 5 meters of cable for the standard setup and three decoupling configurations is shown in figure~\ref{fig:waveforms}. The dual cable setup demonstrates the best performance with the highest amplitude and smallest reflection. The first single cable setup employs a decoupling with a capacitance of 1 nF and a resistance of 10 k$\Omega$ and shows a signal with a large amplitude and reflection. Increasing the resistor value in setup two to 100 k$\Omega$ reduces the reflection while also decreasing the signal amplitude. The third setup includes anode termination of 50 $\Omega$ as an addition to the second setup, which further attenuates the signal, but reduces reflections to a minimum. 

The default resistor ratio values provided by Hamamatsu allows for standard operation of the R11410-10 for intermediate purposes. This ratio can be optimized for environments susceptible to saturation of large light signals. As mentioned previously, nonlinearity at the anode stems from the space charge effect when many electrons become clustered as they travel from the last few dynodes to the anode. Adjusting the resistor ratios at the end of the dynode chain can help offset this effect by increasing the electric field present. 

Figure~\ref{fig:HamamatsuDynodes} shows the results of the anode linearity measurement for several resistor ratio values including the default value of 1:2:1. A 50\% increase in the maximum current which still preserves linearity can be observed by changing the resistor ratio to 1:1.5:1.5. This results in the anode remaining linear to greater than 120 mA. 

\begin{figure}[htb]
\begin{center}
\includegraphics[height = 75mm]{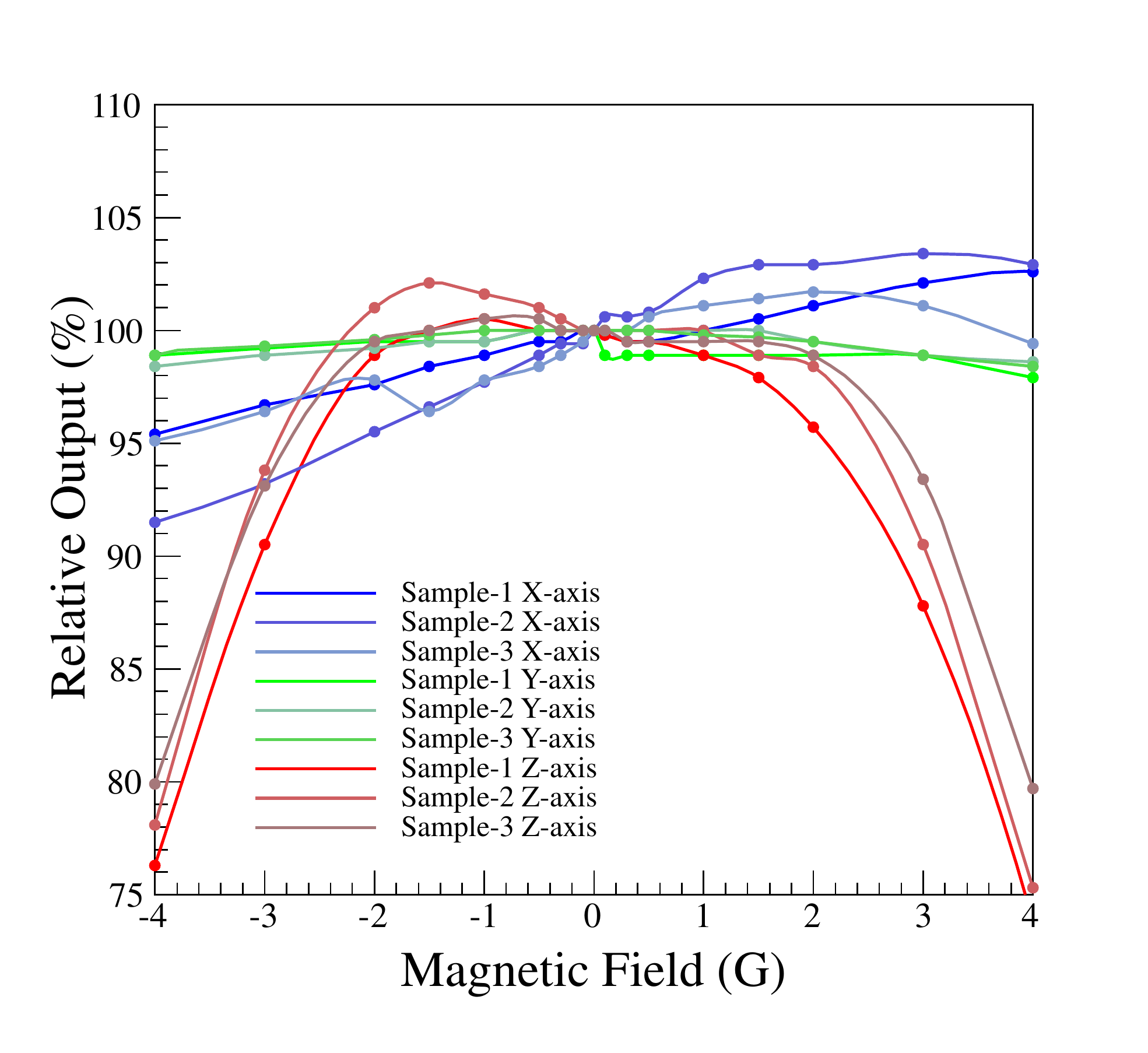}
\end{center}
\caption{
 \label{fig:magneticfield}
 Effect of the magnetic field on the output signal of the R11410-10 PMT. Deviations below 5\% are observed for each orientation of the device except for magnetic fields greater than two gauss.
}
\end{figure}

\subsection{Magnetic Field}
The output signal of the R11410-10 PMT with an external magnetic field applied has been measured for various physical configurations to determine any negative effects associated with large magnetic fields. The external structure of the PMT, an iron-nickel-cobalt alloy called Kovar, has been designed to allow for uniform thermal expansion properties, but it also has a very high magnetic permeability to shield against magnetic fields. This may be applicable for both the Earth's magnetic field ($<$ 1 gauss) as well as any ambient magnetic fields present in the detector environment.

Figure~\ref{fig:magneticfield} shows the properties of the phototube for external applied magnetic fields in any direction. For fields below two gauss, the output signal only deviates by 5\% from a zero-field environment, which demonstrates that the effects of the Earth's magnetic field are negligible.

\section{Summary}
Systematic tests performed at Hamamatsu and UCLA with the R11410-10 PMT have demonstrated its viability for future dark matter experiments. In particular, the high quantum efficiency (above 30\% for units tested), low radioactivity ($\sim$ 20 mBq/piece), high gain (up to $1\times10^{7}$), and low noise properties (50 Hz dark count rate and good single photoelectron resolution) provide the necessary qualities needed for low energy and rare event searches. Additionally, the robust physical characteristics will allow the PMT to maintain high performance over the lifetime of a detector.

\section{Acknowledgements}
We would like to thank Y. Meng, E. Pantic, P. R. Scovell, and H. Wang for their input and advice regarding photomultiplier tube bases and operation. We also thank E. Aprile and C. Galbiati for their encouragement and support. This work was supported in part by DOE grant DE-FG02-91ER40662 and by NSF grant PHY-0919363.

\bibliographystyle{model1-num-names}

\end{document}